\documentclass{revtex4}
\usepackage{amssymb}
\usepackage{amsmath}

\begin{document}
\title{Spectroscopy of Rotating Linear
Dilaton Black Holes From Boxed
Quasinormal Modes }
\author{I. Sakalli}
\email{izzet.sakalli@emu.edu.tr}
\author{G. Tokg\"{o}z}
\email{gulnihal.tokgoz@emu.edu.tr}
\affiliation{Physics Department, Eastern Mediterranean University, Famagusta, Northern
Cyprus, Mersin 10, Turkey}
\date{\today }

\begin{abstract}
Numerical studies of the coupled Einstein-Klein-Gordon system have recently
revealed that confined scalar fields generically collapse to form caged
black holes. In the light of this finding, we analytically study the
characteristic resonance spectra of the confined scalar fields in rotating linear dilaton black
hole geometry. Confining mirrors (cage) are assumed to be placed in
the near-horizon region of a caged rotating linear dilaton black
hole $\frac{r_{m}-r_{2}}{r_{2}}\ll 1$
 ($r_{m}$ is the radius of the cage and $r_{2}$ represents the event
horizon). The radial part of the Klein-Gordon equation is written as a
Schr\"{o}dinger-like wave equation, which reduces to a Bessel differential
equation around the event horizon. Using analytical tools and proper
boundary conditions, we obtain the boxed-quasinormal mode frequencies of the caged rotating linear dilaton black
hole.
Finally, we employ Maggiore's method, which evaluates the
transition frequency in the adiabatic invariant quantity from the
highly damped quasinormal modes, in order to investigate the entropy/area spectra
of the rotating linear dilaton black
hole.

\end{abstract}
\maketitle

\section{Introduction}

Quantum mechanics (QM) and general relativity (GR) fail when confronted
with each other's principles, and are therefore limited in their ability to
describe the Universe. Therefore, one must unify the theories to make them
compatible with one another. The resulting theory is expected to describe
the behavior of the Universe, from leptons and quarks to galaxies. This is
the quantum gravity theory (QGT) \cite{Rovelli}. The onset of the QGT dates
back to the 1970s when Hawking \cite{Hawking74,Hawking75} and Bekenstein 
\cite{Bek72,Bek73,Bek74,Bek74n2,Bek75} amalgamated GR and QM by
using the black hole (BH) as a theoretical arena. However, at present, QGT is under
construction and not completed.

In the spirit of Ehrenfest's adiabatic hypothesis \cite{Ehren,Ehren2},
Bekenstein \cite{Bek74n2} conjectured that the surface area of a BH can only
take on a discrete spectrum of values:

\begin{equation}
A_{n}=\varepsilon n\hbar \text{ \ \ \ \ \ \ }(n=1,2,3,....),  \label{1}
\end{equation}

where $\varepsilon $ is a dimensionless constant. One can easily deduce from Eq. (1) that the minimum change in the horizon area corresponds to $\Delta
A_{\min }=A_{n}-A_{n-1}=\varepsilon \hbar $. Namely, the BH horizon is formed by patches of equal area \cite{Bek74n2}.
Based on Bekenstein's conjecture, many different attempts have been made in the
literature to derive the quantum spectrum of BHs, and consequently different spectra with different $\varepsilon$ have been obtained. The best-known values of $\varepsilon$ are $8\pi $ \cite{Bek74n2,Magg} and $4lnp$ where $p=2,3,...$ \cite{Mukhon1,Mukhon2,HodPRL,Bek2015}. 
Among the methods applied, one of the most important contributions was made by Hod \cite{HodPRL}, who suggested that $%
\varepsilon $ could be determined by utilizing the quasinormal mode (QNM)
frequencies \cite{BertiCQG}. QNMs are the characteristic sounds of BHs and neutron stars \cite{Nollert}. Hod considered the real part of the asymptotic QNM as a transition frequency
(the smallest energy a BH can emit) in the semiclassical limit, and in
sequel showed that its quantum emission gives rise to a change in the mass
of BH, which is related to $A_{n}$. In this way, he found that $\varepsilon =4\ln
3$ (for a Schwarzschild BH) \cite{HodPRL}. Later on, Kunstatter \cite{Kunst} managed to quantize the area of the
Schwarzschild BH with the help of the adiabatic invariance:
\begin{equation}
I_{adb}=\int \frac{dE}{\Delta \omega },  \label{2}
\end{equation}

where $I_{adb}$ is the adiabatic invariant quantity and $\Delta \omega $
denotes the transition frequency between the two neighboring levels of a BH
possessing total energy $dE=T_{H}dS_{BH}$ in which $T_{H}$ and $S_{BH}$
stand for the Hawking temperature and the Bekenstein-Hawking entropy,
respectively. The result obtained was nothing but Hod's result \cite{HodPRL}: the reproduction of $\varepsilon=4\ln 3$.

In 2007, Maggiore \cite{Magg} postulated that a BH can be viewed as a damped
harmonic oscillator whose physically relevant frequency is identical to
the complex QNM frequencies having both real and imaginary parts. Using the
fact that for the highly excited QNMs the imaginary part is dominant over
the real part, he activated Kunstatter's formula (2). According to the Bohr-Sommerfeld quantization rule (BSQR)
 $I_{adb}\simeq n\hbar $ when the quantum number $n\rightarrow
\infty $. Considering this fact, Maggiore proved that for the Schwarzschild
BH the area spectrum is exactly equal to Bekenstein's
original result \cite{Bek74n2}. Following the seminal study of Maggiore, the area
spectra of numerous BHs have been thoroughly investigated (see for instance 
\cite{MMA,MMB,MM1,MM2,MM2b,MM2c,MM3,MM4,MM5,MM6,MM6b,MM7,MM8,MM9}).

In this paper, we study the spectroscopy of a rotating linear dilaton black hole (RLDBH) \cite{Clem2003}, which is a solution to the
Einstein-Maxwell-dilaton-axion (EMDA) theory \cite{Leygnac}. To this end,
we use Maggiore's method (MM) \cite{Magg}. However while doing this computation, instead of the ordinary
QNMs (or vibrational modes) we shall consider the boxed-quasinormal modes (BQNMs) (also known as
quasi-bound states) \cite{Bqnm1}. These modes
reveal when the considered BH is caged \cite{Okaw}. Namely, RLDBH is
assumed to be confined in a cavity with finite-volume. In this setup,
confinement can be achieved by placing an artificial spherical reflecting
surface at some distance $r_{m}$. This configuration is akin to the
\textquotedblleft perfect mirror\textquotedblright\ (used by Press and
Teukolsky \cite{Press}) to perform a BH bomb. However, for studying the resonant
frequencies of the caged BHs, one can follow the recent studies of \cite{Bqnm1,Bqnm2} in which $r_{m}$ was chosen to be close to the near-horizon (NH) region \cite{epjc15}. The propagating scalar fields $\Psi $ are imposed to vanish at $r_{m}$. This
requirement is fulfilled by using two boundary conditions: the Dirichlet
boundary condition (DBC) ($\left. \Psi (r)\right\vert _{r=r_{m}}=0$) and the
Neumann boundary condition (NBC) ($\left. \frac{d\Psi (r)}{dr}\right\vert
_{r=r_{m}}=0$). In addition to this, one should use the fact that QNMs are the pure
ingoing waves at the event horizon. These three conditions guide us to find
the resonance condition, which yields the BQNMs after some manipulations. Here,
we will use the transition frequency of the BQNMs in Eq. (2) to derive the
RLDBH's spectroscopy (area/entropy spectra). It is worth mentioning that the
spectroscopy of RLDBH has recently been studied by Sakalli \cite{MM9}
via the standard QNMs, and the present paper aims to give support to that study.

This paper is arranged as follows. In Sec. 2, we briefly overview the RLDBH
metric, analyze the Klein-Gordon equation (KGE) for a massless scalar field, and derive the Zerilli
equation (ZE) \cite{CS} in this geometry. In Sec. 3, it is shown that the ZE
reduces to a Bessel differential equation around the NH. Next, we impose the
required boundary conditions for computing the complex BQNMs of the caged RLDBH.
Then, applying MM we obtain the spectroscopy of RLDBH. Section
4 concludes . (Throughout the paper we use units in which
fundamental constants are $c=k_{B}=G=1$).

\section{RLDBH and Separation of Massless KGE}

The action of EMDA theory is given by \cite{Leygnac}

\begin{equation}
S=\frac{1}{16\pi }\int d^{4}x\sqrt{\left\vert g\right\vert }\left\{ R-\frac{1%
}{2}e^{4\phi }\partial _{\mu }\aleph \partial ^{\mu }\aleph -2\partial _{\mu
}\phi \partial ^{\mu }\phi \right. \\
-\left. \aleph F_{\mu \nu }\widetilde{F}^{\mu \nu }-e^{-2\phi }F_{\mu \nu
}F^{\mu \nu }\right\} ,  \label{3}
\end{equation}

where $\phi $ and $\aleph $\ denote the dilaton and axion (pseudoscalar
coupled to an Abelian vector field $\mathcal{A}_{em}$) fields, respectively. 
$R$ denotes the Ricci scalar, $F_{\mu \upsilon }$ is the well-known Maxwell
two-form associated with a $U(1)$ subgroup of $E_{8}\times E_{8}$ (or Spin$%
(32)/Z_{2}$), and $\widetilde{F}^{\mu \nu }$ stands for the dual of $F_{\mu
\upsilon }$. RLDBH spacetime is described by \cite{Clem2003,Leygnac}

\begin{equation}
ds^{2}=-f(r)dt^{2}+\frac{dr^{2}}{f(r)}+h(r)\left[ d\theta ^{2}+\sin
^{2}\theta \left( d\varphi -\frac{a}{h(r)}dt\right) ^{2}\right] ,  \label{4}
\end{equation}

where 
\begin{equation}
h(r)=rr_{0},  \label{5}
\end{equation}

\begin{equation}
f(r)=\frac{Z}{h(r)},  \label{6}
\end{equation}

are the metric functions. The constant parameter $r_{0}$ seen in Eq. (5) is
associated with the background electric charge $Q:$ $r_{0}=\sqrt{2}Q$. In Eq. (6), $%
Z=(r-r_{2})(r-r_{1})$, where $r_{1}$ and $r_{2}$ denote the inner and outer\
(event) horizons, respectively, derived from the condition of $f(r)=0$. Explicit form of
these radii are

\begin{eqnarray}
r_{1} &=&M-\sqrt{M^{2}-a^{2}},  \nonumber \\
r_{2} &=&M+\sqrt{M^{2}-a^{2}},  \label{7}
\end{eqnarray}

where the constant $M$ is related to mass, and $a$ is the rotation
parameter modulating the angular momentum: $J=\frac{ar_{0}}{2}$.
For a BH solution, $M\geq a$. The non-asymptotically flat geometry of the RLDBH entails
the computation of the quasilocal mass $M_{QL}$\ \cite{BYform}. As a result,
one finds that $M=2M_{QL}$. Furthermore, the background fields are given by

\begin{equation}
e^{-2\phi }=\frac{h(r)}{s(r)},  \label{8}
\end{equation}

\begin{equation}
\aleph =-\frac{r_{0}a\cos \theta }{s(r)},  \label{9}
\end{equation}

where $s(r)=r^{2}+a^{2}\cos ^{2}\theta $. The electromagnetic four-vector
potential reads

\begin{equation}
\mathcal{A}_{em}\mathcal{=}\frac{1}{\sqrt{2}}(e^{2\phi }dt+a\sin ^{2}\theta
d\varphi ).  \label{10}
\end{equation}

The Hawking temperature of the RLDBH is provided by the surface gravity $\kappa $
\cite{Wald}\ as

\begin{align}
T_{H}& =\frac{\hbar \kappa }{2\pi }=\frac{\hbar }{2\pi }\left( \left. \frac{%
\partial _{r}f(r)}{2}\right\vert _{r=r_{2}}\right) ,  \nonumber \\
& =\frac{\hbar \left( r_{2}-r_{1}\right) }{4\pi r_{2}r_{0}}.  \label{11}
\end{align}

As can be seen from Eq. (11), in the static case ($a=0$, i.e., $r_{1}=0$) this BH
emits radiation with a constant (mass-independent) temperature, which corresponds to the
isothermal process \cite{PasaIz,AstrSak} leading to the uninformed Hawking radiation \cite{AlSak} . If $A_{H}$ stands for the area of the event horizon, then the RLDBH's entropy is given by

\begin{equation}
S_{BH}=\frac{A_{H}}{4\hbar }=\frac{\pi r_{2}r_{0}}{\hbar }.  \label{12}
\end{equation}

Angular velocity of the RLDBH:

\begin{equation}
\Omega _{H}=-\left. \frac{g_{tt}}{g_{t\varphi }}\right\vert _{r=r_{2}}=\frac{%
a}{r_{2}r_{0}}.  \label{13}
\end{equation}

Thus, the first law of thermodynamics can be validated for the RLDBH via

\begin{equation}
dM_{QL}=T_{H}dS_{BH}+\Omega _{H}dJ.  \label{14}
\end{equation}

Meanwhile, one may examine why Eq. (14) does not contain electric charge $Q$%
. This is because here $Q$ is nothing but a background charge of fixed value \cite%
{Clem2003}.

Utilizing the massless KGE:

\begin{equation}
\partial _{k}(\sqrt{-g}\partial ^{k}\Phi )=0,\text{ \ \ \ }k=0...3,
\label{15}
\end{equation}

with the following ansatz for the scalar field $\Phi $

\begin{equation}
\Phi =\frac{\rho (r)}{\sqrt{r}}e^{-i\omega t}Y_{l}^{m}(\theta ,\varphi ),%
\text{ \ \ }Re(\omega )>0,  \label{16}
\end{equation}

where $Y_{l}^{m}(\theta ,\varphi )$ denotes the spheroidal harmonics with
the eigenvalue $-l(l+1)$ \cite{Damora}, we get

\begin{equation}
Z\partial _{rr}\rho (r)+(\frac{r^{2}-r_{1}r_{2}}{r})\partial _{r}\rho (r)+%
\left[ \frac{\left( \omega rr_{0}-ma\right) ^{2}}{Z}-l(l+1)\right. 
+\left.\frac{(3r_{1}-r)r_{2}-rr_{1}-r^{2}}{4r^{2}}\right] \rho (r)=0,
\label{17}
\end{equation}

From Eq. (17), one can obtain the ZE \cite{CS} as follows

\begin{equation}
\left( -\partial _{r^{\ast }r^{\ast }}+V\right) \rho (r^{\ast })=\omega
^{2}\rho (r^{\ast }),  \label{18}
\end{equation}

where the effective potential or the so-called Zerilli potential is given by

\begin{equation}
V=f(r)\left\{{\frac{1}{h(r)}\left[ \frac{r^{2}+2Mr-3a^{2}}{4r^{2}}+l(l+1)%
\right]}-\frac{m\widetilde{\Omega }}{f(r)}(m\widetilde{\Omega 
}-2\omega )\right\},  \label{19}
\end{equation}

by which

\begin{equation}
\widetilde{\Omega }=\frac{a}{rr_{0}}.  \label{20}
\end{equation}

The tortoise coordinate $r^{\ast }$ is defined as

\begin{equation}
r^{\ast }=\int \frac{dr}{f(r)},  \label{21}
\end{equation}

which yields

\begin{equation}
r^{\ast }=\frac{r_{0}}{r_{2}-r_{1}}\ln \left[ \frac{(\frac{r}{r_{2}}%
-1)^{r_{2}}}{(r-r_{1})^{r_{1}}}\right] .  \label{22}
\end{equation}

The asymptotic limits of $r^{\ast }$ become

\begin{equation}
\lim_{r\rightarrow r_{2}}r^{\ast }=-\infty \text{ \ \ and \ }%
\lim_{r\rightarrow \infty }r^{\ast }=\infty .  \label{23}
\end{equation}

\section{BQNM Frequencies of Caged RLDBH and Spectroscopy Computation}

In this section, we aim to read the BQNM frequencies of the caged RLDBH. These
resonant frequencies can be analytically determined when the
confining mirrors are placed in the vicinity of the event horizon. To this
end, we will impose the Hod's boundary condition that nothing
is supposed to come out of the event horizon of the caged BH and to survive at the
cage  \cite{Bqnm1} .

Setting

\begin{equation}
y=\frac{r-r_{2}}{r_{2}},  \label{24}
\end{equation}

the metric function $f(r)$ transforms into

\begin{equation}
f(r)\rightarrow f(y)=\frac{1}{r_{0}}\left( r_{2}y-\frac{y}{1+y}r_{1}\right) ,
\label{25}
\end{equation}

which has the following NH behavior:

\begin{equation}
f_{NH}(y)\cong \tau y+O(y^{2}),  \label{26}
\end{equation}

wherein

\begin{equation}
\tau =\frac{r_{2}-r_{1}}{r_{0}}.  \label{27}
\end{equation}

Thus, the tortoise coordinate (21) around the NH behaves as

\begin{equation}
r^{\ast }\cong \int \frac{r_{2}dy}{f_{NH}(y)}\cong \frac{r_{2}}{\tau }\ln y=%
\frac{1}{2\kappa }\ln y,  \label{28}
\end{equation}

so that we have

\begin{equation}
y=e^{2\kappa r^{\ast }}.  \label{29}
\end{equation}

Substituting Eq. (24) into Eq. (19), one finds the NH Zerilli potential
as follows

\begin{equation}
V_{NH}(y)=\alpha +Fy+O(y^{2}),  \label{30}
\end{equation}

where the parameters of Eq. (30)\ are\ given by

\begin{equation}
\alpha =m\Omega _{H}\left( \widetilde{\omega }+\omega \right) ,  \label{31}
\end{equation}

\begin{equation}
F=\frac{\kappa }{2r_{0}r_{2}^{2}}\left\{ \left[ \left( 2l+1\right) r_{2}%
\right] ^{2}+2Mr_{2}-3a^{2}\right\} -2m\Omega _{H}\widetilde{\omega },
\label{32}
\end{equation}

in which

\begin{equation}
\widetilde{\omega }=\omega -m\Omega _{H}.  \label{33}
\end{equation}

Here $\widetilde{\omega }$ connotes the wave frequency measured by the
observer rotating with the horizon \cite{Clem2003}. Therefore, the ZE
(18) around the NH becomes

\begin{equation}
\left[ -\partial _{r^{\ast }r^{\ast }}+\alpha +Fe^{2\kappa r^{\ast }}\right]
\rho (r^{\ast })=\omega ^{2}\rho (r^{\ast }),  \label{34}
\end{equation}

One can find that the general solution of Eq. (34) is given by

\begin{equation}
\rho (r^{\ast })=D_{1}J_{-i\overline{\omega }}(u)+D_{2}Y_{-i\overline{\omega 
}}(u),  \label{35}
\end{equation}

where $u=2i\sqrt{\eta }e^{\kappa r^{\ast }}$ and $D_{1},_{2}$ are constants. 
$J_{-i\overline{\omega }}(u)$ and $Y_{-i\overline{\omega }}(u)$ denote the
Bessel functions \cite{AS} of the first and second kinds, respectively. The
parameters of the Bessel functions are given by%
\begin{equation}
\eta =\frac{Fr_{2}^{2}}{\tau ^{2}},  \label{36}
\end{equation}

\begin{equation}
\overline{\omega }=\frac{\widetilde{\omega }}{\kappa }.  \label{37}
\end{equation}

Using the following limiting forms (when $z\rightarrow 0$) of the Bessel
functions \cite{AS,OR}

\begin{equation}
J_{p}(z)\sim \frac{\left( \frac{1}{2}z\right) ^{p}}{\Gamma (1+p)},\text{ \ \
\ \ \ \ \ }\left( p\neq -1,-2,-3,.....\right) ,  \label{38}
\end{equation}

\begin{equation}
Y_{p}(z)\sim -\frac{\Gamma (p)}{\pi }\left( \frac{1}{2}z\right) ^{-p},\text{
\ }(\Re p>0),  \label{39}
\end{equation}

we get the NH ($e^{\kappa r^{\ast }}\ll 1$) behavior of Eq. (35):

\begin{equation}
\rho \sim D_{1}\frac{\left( i\sqrt{\eta }\right) ^{-i\overline{\omega }}}{%
\Gamma (1-i\overline{\omega })}e^{-i\widetilde{\omega }r^{\ast }}-D_{2}\frac{%
1}{\pi }\Gamma (-i\overline{\omega })\left( i\sqrt{\eta }\right) ^{i%
\overline{\omega }}e^{i\widetilde{\omega }r^{\ast }}.  \label{40}
\end{equation}

One can deduce that $D_{1}$ and $D_{2}$ are related to the amplitude of the
ingoing and outgoing waves at NH, respectively. Now, one can impose the
condition of having QNMs, which stipulates purely ingoing waves at the horizon. Hence, we
pick $D_{1}\neq 0$ and set $D_{2}=0$. Thus, the physical solution of Eq.
(35) is given by

\begin{equation}
\rho (r^{\ast })=D_{1}J_{_{-i\overline{\omega }}}(u),  \label{41}
\end{equation}

or

\begin{equation}
\rho (y)=D_{1}J_{-i\overline{\omega }}(2i\sqrt{\eta y}).  \label{42}
\end{equation}

Following \cite{Bqnm1,Bqnm2,Okaw}, we consider the DBC at the surface $%
y\equiv y_{m}=\frac{r_{m}-r_{2}}{r_{2}}$\ of the confining cage:

\begin{equation}
\left. \rho (y)\right\vert _{y=y_{m}}=J_{-i\overline{\omega }}(2i\sqrt{\eta
y_{m}})=0.  \label{43}
\end{equation}

Using the following relation \cite{AS}

\begin{equation}
Y_{p}(z)=J_{p}(z)\cot (p\pi )-J_{-p}(z)\csc (p\pi ),  \label{44}
\end{equation}

we can rewrite the condition (43) as

\begin{equation}
\tan (i\pi \overline{\omega })=\frac{J_{i\overline{\omega }}(2i\sqrt{\eta
y_{m}})}{Y_{i\overline{\omega }}(2i\sqrt{\eta y_{m}})},  \label{45}
\end{equation}

which is called the resonance condition \cite{Bqnm1,Bqnm2}. According to Hod \cite{Bqnm1}, the boundary of the confining cavity leading to the characteristic resonance spectra is close to the NH ($r_{m}\approx r_{2}$). So

\begin{equation}
l_{m}\equiv \eta y_{m}\ll 1.  \label{46}
\end{equation}

With the aid of Eqs. (38) and (39), we find the resonance condition (45) approximates to

\begin{eqnarray}
\tan (i\overline{\omega }\pi ) &\sim &-\frac{\pi \left( i\sqrt{l_{m}}\right)
^{2i\overline{\omega }}}{\Gamma \left( i\overline{\omega }+1\right) \Gamma
\left( i\overline{\omega }\right) },  \nonumber \\
&=&i\frac{\pi e^{-\pi \overline{\omega }}}{\overline{\omega }\Gamma
^{2}\left( i\overline{\omega }\right) }\left( l_{m}\right) ^{i\overline{%
\omega }}.  \label{47}
\end{eqnarray}

Finally, one should also impose the NBC \cite{Bqnm1} on the solution (42):

\begin{equation}
\left. \frac{d\rho (y)}{dy}\right\vert _{y=y_{m}}=0,  \label{48}
\end{equation}

which admits the following expression \cite{AS}

\begin{equation}
J_{-i\overline{\omega }-1}(2i\sqrt{l_{m}})-J_{-i\overline{\omega }+1}(2i%
\sqrt{l_{m}})=0.  \label{49}
\end{equation}

Recalling Eq. (44), one can reproduce the following expression

\begin{eqnarray}
Y_{p+1}(z)-Y_{p-1}(z) &=&\cot (p\pi )\left[ J_{p+1}(z)-J_{p-1}(z)\right] 
\nonumber \\
&&-\csc (p\pi )\left[ J_{-p-1}(z)-J_{-p+1}(z)\right] .  \label{50}
\end{eqnarray}

Substituting Eq. (49) into Eq. (50), we derive the resonance condition for the NBC:

\begin{equation}
\tan (i\pi \overline{\omega })=\frac{J_{i\overline{\omega }-1}(2i\sqrt{l_{m}}%
)}{Y_{i\overline{\omega }+1}(2i\sqrt{l_{m}})}\left[ \frac{-1+\frac{J_{i%
\overline{\omega }+1}(2i\sqrt{l_{m}})}{J_{i\overline{\omega }-1}(2i\sqrt{%
l_{m}})}}{1-\frac{Y_{i\overline{\omega }-1}(2i\sqrt{l_{m}})}{Y_{i\overline{%
\omega }+1}(2i\sqrt{l_{m}})}}\right] .  \label{51}
\end{equation}

From Eqs. (38) and (39), we get

\begin{equation}
\frac{Y_{i\overline{\omega }-1}(2i\sqrt{l_{m}})}{Y_{i\overline{\omega }+1}(2i%
\sqrt{l_{m}})}\equiv \frac{J_{i\overline{\omega }+1}(2i\sqrt{l_{m}})}{J_{i%
\overline{\omega }-1}(2i\sqrt{l_{m}})}\sim O\left( l_{m}\right) ,  \label{52}
\end{equation}

Thus, we can rewrite the resonance condition (51) as

\begin{eqnarray}
\tan (i\pi \overline{\omega }) &\sim &-\frac{J_{i\overline{\omega }-1}(2i%
\sqrt{l_{m}})}{Y_{i\overline{\omega }+1}(2i\sqrt{l_{m}})},  \nonumber \\
&=&-i\frac{\pi e^{-\pi \overline{\omega }}}{\overline{\omega }\Gamma
^{2}\left( i\overline{\omega }\right) }\left( l_{m}\right) ^{i\overline{%
\omega }}.  \label{53}
\end{eqnarray}

According to Eq. (46), the damped modes with ${Im}(\overline{\omega })<0
$ lead to $\left( l_{m}\right) ^{i\overline{\omega }}\ll 1$. So, it is clear
that both resonance conditions (47) and (53) take small quantities.

We now use an iteration scheme to obtain the BQNMs of the caged RLDBH. For both DBC and NBC, the $%
0^{th} $ order resonance condition is given by \cite{Bqnm1}:

\begin{equation}
\tan (i\pi \overline{\omega }_{n}^{(0)})=0,  \label{54}
\end{equation}

which results in

\begin{equation}
\overline{\omega }_{n}^{(0)}=-in,\text{ \ \ \ \ \ \ \ \ \ }(n=1,2,3,....).
\label{55}
\end{equation}

The $1^{st}$ order resonance condition is obtained after substituting Eq.
(55) into r.h.s of Eqs. (47) and (53). Thus, we have

\begin{equation}
\tan (i\overline{\omega }_{n}^{(1)}\pi )=\pm i\frac{\pi e^{i\pi n}}{\left(
-in\right) \Gamma ^{2}\left( n\right) }\left( l_{m}\right) ^{n},  \label{56}
\end{equation}

which is equivalent to

\begin{equation}
\tan (i\overline{\omega }_{n}^{(1)}\pi )=\mp n\frac{\pi \left( -l_{m}\right)
^{n}}{\left( n!\right) ^{2}},  \label{57}
\end{equation}

where plus (minus) stands for the NBC (DBC). In the $u\ll 1$ regime, we have

\begin{equation}
\tan (u+n\pi )=\tan (u)\approx u.  \label{58}
\end{equation}

Using this for Eq. (57), we find the characteristic resonance spectra as
follows

\begin{equation}
i\overline{\omega }_{n}\pi =n\pi \mp n\frac{\pi \left( -l_{m}\right) ^{n}}{%
\left( n!\right) ^{2}},  \label{59}
\end{equation}

which yields

\begin{equation}
\overline{\omega }_{n}=-in\left[ 1\mp \frac{\left( -l_{m}\right) ^{n}}{%
\left( n!\right) ^{2}}\right] .  \label{60}
\end{equation}

Putting this into Eqs. (37) and (33) gives the BQNMs of the RLDBH

\begin{equation}
\omega _{n}=m\Omega _{H}-i\kappa n\left[ 1\mp \frac{\left( -l_{m}\right) ^{n}%
}{\left( n!\right) ^{2}}\right] ,\text{ \ \ \ \ }(n=1,2,3,....).
\label{61}
\end{equation}

For the highly damped modes, Eq. (61) becomes

\begin{equation}
\omega _{n}\approx m\Omega _{H}-i\kappa n,\text{ \ \ \ \ \ \ \ \ \ \ \ \ (}%
n\rightarrow \infty \text{),}  \label{62}
\end{equation}

which is in accordance with the recent study of Sakalli \cite{MM9} in which
the quantization of the RLDBH was studied with the standard QNMs. According
to MM, the transition frequency now reads

\begin{equation}
\Delta \omega \approx {Im}(\Delta \omega )={Im}(\omega _{n}-\omega
_{n-1})=\kappa =\frac{2\pi T_{H}}{\hbar }.  \label{63n}
\end{equation}

From Eqs. (2) and (63), we now have

\begin{equation}
I_{adb}=\frac{\hbar }{2\pi }S_{BH}.  \label{64}
\end{equation}

Acting upon the BSQR ($I_{adb}=\hbar n$), we obtain the entropy spectrum as

\begin{equation}
S_{BH\_n}=2\pi n,  \label{65}
\end{equation}

which also yields the area spectrum as follows

\begin{equation}
A_{n}=4\hbar S_{BH\_n}=8\pi \hbar n.  \label{66}
\end{equation}

Thus, the minimum area spacing of the RLDBH becomes

\begin{equation}
\Delta A_{\min }=8\pi \hbar .  \label{67}
\end{equation}

The above results support Bekenstein's conjecture \cite{Bek74n2}.
Furthermore, we see that the spectroscopy of the RLDBH is independent from its
characteristic parameters, and its area spectrum is evenly spaced.

\section{Conclusion}
Motivated by recent studies \cite{Bqnm1,Bqnm2} which have shown that the BQNMs occur when a BH is caged by a perfect mirror (artificial) located in the NH region, we have studied the entropy/area spectra of RLDBHs. To this end, we first analyzed the KGE for a massless scalar field in this geometry. Using the
separability feature of the KGE, we derived the ZE (18) in the
background of the RLDBH, and read its effective potential (19). It has been
shown that NH-ZE (34) has a solution in terms of the Bessel functions (35). Thus, by using
the boundary conditions of QNM (only ingoing waves survive at the NH), DBC, and
NBC, we obtained the BQNMs (61) analytically. Then, the transition
frequency obtained from the highly damped BQNMs was substituted in Eq. (2) to compute the quantum entropy/area spectra of RLDBH. The resulting area spectrum is equidistant and fully consistent with Bekenstein's conjecture \cite{Bek74n2}. Finally, our results support the study of Sakalli \cite{MM9}, which proves that RLDBH entropy/area spectra  are equally spaced from ordinary QNMs.

\section*{Acknowledgments}

We would like to thank G\'{e}rard Cl\'{e}ment, Shahar Hod, and Mustafa Halilsoy for helpful discussions.

\end{document}